\documentclass[11pt,a4paper,english,amssymb,nofootinbib,superscriptaddress]{revtex4}
\usepackage{lmodern}

\usepackage[T1]{fontenc}
\usepackage[latin9]{inputenc}
\usepackage{babel}
\usepackage{amsmath}
\usepackage{graphicx}  
\usepackage{amssymb}
\usepackage{esint}
\usepackage[unicode=true, pdfusetitle,
 bookmarks=true,bookmarksnumbered=false,bookmarksopen=false,
 breaklinks=false,pdfborder={0 0 1},backref=false,colorlinks=false]
 {hyperref}
\def\b{\begin{equation}}
\def\e{\end{equation}}

\makeatletter
\@ifundefined{textcolor}{}
{%
 \definecolor{BLACK}{gray}{0}
 \definecolor{WHITE}{gray}{1}
 \definecolor{RED}{rgb}{1,0,0}
 \definecolor{GREEN}{rgb}{0,1,0}
 \definecolor{BLUE}{rgb}{0,0,1}
 \definecolor{CYAN}{cmyk}{1,0,0,0}
 \definecolor{MAGENTA}{cmyk}{0,1,0,0}
 \definecolor{YELLOW}{cmyk}{0,0,1,0}
 }

\usepackage{latexsym}\usepackage{bm}

\makeatother

\makeatother

\begin{document}

\title{More on Cotton Flow }

\author{Ercan Kilicarslan}

\email{kercan@metu.edu.tr}

\affiliation{Department of Physics,\\
 Middle East Technical University, 06800, Ankara, Turkey}

\author{Suat Dengiz}

\email{suat.dengiz@metu.edu.tr}

\affiliation{Department of Physics,\\
 Middle East Technical University, 06800, Ankara, Turkey}

\author{Bayram Tekin}

\email{btekin@metu.edu.tr}

\affiliation{Department of Physics,\\
 Middle East Technical University, 06800, Ankara, Turkey}

\date{\today}

\begin{abstract}
Cotton flow tends to evolve a given initial metric on a three manifold to a conformally flat one. Here we expound upon the earlier work on Cotton flow and study the linearized version of it around a generic initial metric by employing a modified form of the DeTurck trick. We show that the flow around the flat space, as a critical point, reduces to an anisotropic generalization of linearized KdV equation with complex dispersion relations one of which is an unstable mode, rendering the flat space unstable under small perturbations. We also show that Einstein spaces and some conformally flat non-Einstein spaces are linearly unstable. We refine the gradient flow formalism and compute the second variation of the entropy and show that generic critical points are extended Cotton solitons. We study some properties of these solutions and find a Topologically Massive soliton that is built from Cotton and Ricci solitons.  In the Lorentzian signature, we also show that the  $pp$-wave metrics are both Cotton and Ricci solitons.  
\end{abstract}
\maketitle

\section{Introduction}

Gravity has a smoothing effect on a mass or matter distribution, as is evident from all the sufficiently massive spherical objects in the universe, including our planet which has deviations from surface smoothness with its tallest mountain and deepest point in the ocean amounting to less than $10^{-3}$ of its radius. For more massive compact objects, the deviations from surface smoothness would be even less: For example on a neutron star with 1.4 times the mass of the Sun, the height of a mountain could be around 1 cm (not much to hike!). Black holes are even smoother: They have "no hair".  Rotations about an axis can introduce larger deviations from perfect spherical shape, but that is a totally different story. For the earth, this deviation is about $4\times10^{-3}$, that is the polar radius is around 25 km smaller than the equatorial one. While gravity's smoothing effect on objects that we can see is clear, what we also know is that space-time, even if devoid of any matter, as a manifold itself, also gravitates due to the non-linearity of the field equations. This raises many questions that fall in the intersection of physics and mathematics. For example, one might wonder if one can study the topological classification of manifolds with the help of gravitational equations of some sort. The answer is affirmative but not obvious and actual execution of such a classification is certainly not straightforward. First, one needs to make sure that there is a connection between the Riemannian structure (the metric, the curvature) and the topology. After all the metric and the curvature are local quantities but global topology a priori should not be determined by the local structure. This cursory look is actually not correct and there is a connection between the topology of a manifold and its Riemannian structure as beautifully exemplified by the formula relating the Euler number of a closed two surface in terms of its total Gaussian curvature: $2 \pi\chi (\Sigma)= \int_\Sigma K \sqrt{g} \, d^2 x .$

This and analogous results in higher dimensions encourage one to study the global topology of manifolds with the help of curvature and the metric. Considering the above mentioned smoothing nature of gravity, and the fact that topological classification boils down to identifying manifolds that can be smoothly deformed to each other, one might perhaps try to let the metric on a manifold evolve in time with Einstein's equation or some modified form of it. While this seems like a feasible idea, it is also well-known that singularities, such as black hole singularity, form in a sufficiently general non-linear theory of gravity. Hence one must somehow deal with singularities or better yet turn them into advantage as is done in Ricci flow with surgery.

The choice of gravity-like equations in order to study the topology of manifolds is an important issue. For example, if one naively takes the Einstein's equations (say in vacuum), then depending on the number of dimensions, one has a very different nature of solutions. Considering three manifolds \emph{per se}, one can split the four dimensional Einstein's equations into $3+1$ as space and time hence study the evolution of $3$ spaces in time. But it turns out that there are all sorts of solutions, including propagating local modes (gravitons), gravitational waves etc. due to the hyperbolic nature of the equations. Sticking to three dimensional manifolds, the goal is to have a finite dimensional solution space (up to perhaps scaling of the metric and the diffeomorphisms). In the choice of the proper equations, uniformization theorem of two dimensional closed, orientable surfaces, which reduces to finding conformally equivalent metrics of constant Gaussian curvature (usually normalized as $1, 0, -1$ for the sphere, torus and higher genus tori respectively), gives us an important hint: That is, one must find "constant curvature" metrics on a given manifold. Here  by "constant-curvature", we mean either constant scalar curvature or constant Ricci curvature or in the extreme case, constant Riemann curvature. Of course the last one is too restrictive and its solutions are already known; they are the maximally-symmetric spaces. On the other hand, finding constant Ricci curvature metrics boils down to solving Einstein's equations in vacuum ($R_{ij}=\frac{R}{n} g_{ij}$) for which a general strategy is not known even though there are books compiling plethora of solutions. The question whether a manifold of dimension larger than two admits a constant scalar curvature is called the Yamabe problem \cite{yamabe}, which in some sense generalizes the two dimensional uniformization problem to higher dimensions.

All the above discussion suggests that given a differentiable manifold ${\cal M}$ and an initial metric $g_0$ on it, we should try to get a "nicer", more symmetric metric by smoothly deforming the initial metric. The way to proceed is clear: As in the case of the usual diffusive scalar heat equation, one must have an equation of the form
\begin{equation}
\frac{\partial}{\partial t} g_{ij}(t, {x})=  E_{ij} (t, {x}),
\end{equation}
where the deformation parameter $t$ is "time" (or a parameter related to the energy scale) and it is an external parameter not a dimension on the manifold, while $x$ denotes the local coordinates on the manifold which we shall take to have a positive signature metric. Depending on the property of the manifold one is interested in, a specific $E_{ij}$ tensor can be chosen. For example, to define the Ricci flow, which was introduced by Hamilton \cite{Hamilton1}, one takes the following "volume-normalized " equation 
   \begin{equation}
   \frac{\partial}{\partial t} g_{ij}=-2 \Big (R_{ij}-\frac{r g_{ij}}{n} \Big),
   \end{equation}
where $r$ is the average Ricci scalar
\begin{equation}
r=\mbox{V}^{-1} \int  d^n x \sqrt{g} \,R ,\hskip 0.6 cm \mbox{where} \hskip 0.6 cm \mbox{V}\equiv \int  d^n x\,\sqrt{g} \, .
\end{equation}
Fixed points of the flow are Einstein metrics, therefore this is in some sense a parabolic extension of Einstein's equation, albeit its parabolic nature is not apparent at this non-linear level. Namely with this flow, the metric is deformed in such a diffusive way that the manifold eventually admits a constant Ricci curvature while preserving volume. Hamilton \cite{Hamilton2} used this equation to give another proof of Poincare's uniformization theorem for two dimensional closed surfaces mentioned above. Over a decade ago, Perelman \cite{Perelman1, Perelman2}, using a modified form of the Ricci flow equation, proved the three dimensional uniformization theorem, then called the Thurston's conjecture \cite{Thurston} with a heroic effort bringing Hamilton's expectation and programme to a successful end. Even though the uniformization theorem itself and the equations employed to prove it are so closely related to physics, not much work has been done on Ricci flow in physics (save a
  couple which we mention below).
 
Let us expound upon how the uniformization theorem and Ricci flow type equations are related to physics. First of all, just about the same time (and actually, a little before) Hamilton introduced the Ricci flow, the equation was found by Friedan \cite{Friedan} as the renormalization group (RG) equations of the couplings in a non-linear two dimensional sigma model. In the non-linear sigma model, the metric is just like any other coupling and runs with energy, hence there is a $\beta$ function for it and the $\beta$ function is given as the Ricci tensor plus infinitely many two tensors built from the curvature tensor and its derivatives. Two-dimensional sigma model fields are coordinates on the target space which is a Riemannian manifold. RG flow in the sigma model corresponds to a modified Ricci flow in the target space.
 
One can also see the relevance of the three dimensional uniformization theorem in general relativity (GR): In the $3+1$ dimensional formulation of GR, three manifolds evolve in time. Since closed 3 manifolds are classified by the uniformization theorem, in essence, the theorem is at the heart of the question of what the shape of the apparent 3 dimensional spatial universe is. Let us give another example of how these flows can be relevant to physics. Consider a gravity theory (a Euclidean one) defined with certain boundary conditions. There could be many solutions satisfying the boundary conditions. In that case, the question is which solution is the \emph{global} minimum of the Euclidean action. One can use a Ricci flow type equation to study the stability of the solutions and determine possible transitions between solutions. In fact, this was done in \cite{Headrick} for general relativity in a box with a boundary of $S^2 \times S^1$ with 3 critical points in four dimensions.
  Their exercise is a 
demonstration of how Euclidean version of gravity (which could possibly correspond to quantum processes such as black hole decay in the Lorentzian setting) can be studied by geometric flows. Namely, the discrete solution space is extended by the Ricci flow hence an off-shell theory can be constructed. See \cite{maloney} for a similar analysis using the Ricci-Cotton flow that studies the transition between the two vacua Anti de Sitter (AdS) and warped AdS. See \cite{Woolgar}, for some other possible applications of Ricci flow and \cite{gegenberg} for a modified Ricci flow with additional fields and \cite{Das} for some higher derivative flows. 
 
Inspired by the work on Ricci flow, in \cite{Cottonflow}, Cotton tensor ($C_{ij}$) was used to define a geometric flow exclusive to three dimensions as  
\begin{equation}
\frac{\partial}{\partial t} g_{ij}(t, {x})= \kappa C_{ij} (t, {x}),
\label{cottonflowaction}
\end{equation}
where fixed points are locally conformally flat metrics. Note that since the Weyl tensor vanishes identically in three dimensions, a manifold is conformally flat if and only if $C_{ij}=0$. Therefore, Cotton flow equation could be used to answer the question: whether a given three manifold ${\cal M}$ admits a conformally flat metric or not. In this sense pure Cotton flow is orthogonal to the Yamabe flow since Yamabe flow keeps the conformal class intact, while the Cotton flow changes the conformal class. In \cite{Cottonflow}, equation (\ref{cottonflowaction}) was derived as a gradient flow of an entropy functional and the equation was numerically and analytically  solved for homogeneous Thurston's geometries. In particular, it was shown that a deformed initial metric on $S^3$ flows into the conformally flat round metric on $S^3$. In the table below, we compile these homogeneous geometries and compute their Cotton tensors just for the sake of completeness. As it is clear from the table, the first five of the geometries are fixed points of the Cotton flow. Moreover, homogeneously deformed initial metrics on $R^3, \,\, H^3, \,\, S^2 \times R^1$ and $H^2 \times R^1$ are left intact under the Cotton flow. On the other hand, under the Ricci flow $S^2 \times R^1$ and $H^2 \times R^1$ geometries are degenerated \cite{Isenberg}. Analytic solutions show that homogeneous metrics in the $Nil$ class tend to a pancake degeneracy and the $Solv$ metrics develop cigar degeneracies. 
\begin{center}
\resizebox{\columnwidth}{!}{%
    \begin{tabular}{ | l | l | l | l| l | p{5 cm}|}
    \multicolumn{4}{c}{Thurston Geometries} \\
\cline{1-4}
    Geometry & Metric & Curvature Tensors & Cotton Tensors \\ \hline
    $R^3$ & $ds^2=dx^2+dy^2+dz^2$&$ R_{ij}=0$& $C_{ij}=0$\\ \hline
   $S^3$ & $ds^2=dx^2+\sin^2 x dy^2+ (dz+ \cos x\, dy)^2$& $R_{ij}=\frac{1}{2}g_{ij}$ & $C_{ij}=0$\\ \hline
   $H^3$ & $ds^2=\frac{1}{x^2} (dx^2+dy^2+dz^2)$& $ R_{ij}=-2 g_{ij}  $ & $C_{ij}=0$ \\ \hline
   $S^2 \times R^1$ & $ds^2=dx^2+\sin^2 x \, dy^2+dz^2$ &$ R_{11}=1, \, R_{22}=\sin^2 x $ &  $C_{ij}=0$ \\ \hline
   $H^2 \times R^1$ & $ds^2=\frac{1}{x^2} (dx^2+dy^2)+dz^2$ &$ R_{11}=R_{22}=-\frac{1}{x^2} $  &  $C_{ij}=0$ \\ \hline
   $Sol$ & $ds^2=e^{2 z}dx^2+e^{-2 z}dy^2+dz^2$ & $R_{33}=-2$ & $ C_{12}=2 $  \\ \hline
   $Nil$ & $ds^2=dx^2+dy^2+(dz+x \,dy )^2$ & $R_{11}=-\frac{1}{2}, \, R_{22}=\frac{x^2-1}{2} $ & $C_{11}=-\frac{1}{2}, C_{22}=x^2-\frac{1}{2}$ \\
   & & $R_{33}=\frac{1}{2},\, R_{23}=\frac{t}{2}$ & $C_{33}=1, \, C_{23}=x $ \\ \hline
   $SL(2, \mathbb{R})$ & $ds^2=\frac{1}{x^2}  (dx^2+dy^2)+(dz+\frac{1}{x} d y )^2$ & $ R^2_{ij}=\frac{19}{4}, \,\,R=-\frac{5}{2},\,R_{11}=-\frac{3}{2x^2},\,$  & $C_{11}=\frac{1}{x^2},\, C_{22}=-\frac{1}{x^2} $ \\ 
   & & $R_{22}=-\frac{1}{x^2},\, R_{23}=\frac{1}{2x},R_{33}=\frac{1}{2}$  & $ C_{23}=-\frac{2}{x},\, C_{33}=-2$\\
   \hline
    \end{tabular}%
    }
\end{center}
\vskip 0.7 cm 
In this current work, we shall study several pertinent points in Cotton flow, the most important being the linear (in)stability of flat space. We also show the instability of Einstein spaces and some conformally flat spaces. We shall also linearize the Cotton flow equations about a generic background and apply DeTurck's technique \cite{DeTurck} to get a slightly better equation, yet we shall not be able to prove short-time existence for an arbitrary initial metric. Nevertheless, ensuing discussions along this line will need our linearized equations. We also refine the earlier gradient flow formulation of the Cotton flow and define new objects which we shall call  extended Cotton solitons that are critical points of the entropy functional. We study two examples of Cotton and Ricci solitons and define Topologically Massive solitons. There has been several other interesting works in the literature regarding the Cotton flow (or its more complicated cousin Ricci-Cotton flow). For example, see \cite{Bakas1, Bakas2, Cartas} for Cotton-like flow in the Horava-Lifshitz gravity context. See also \cite{Cottonsoliton1, Cottonsoliton2, Cottonsoliton33} where Cotton solitons were found.

The lay-out of the paper is as follows: In Sec.II, we linearize the Cotton flow equation about a generic background and identify the symmetries of the equation. In Sec.III, we fix the diffeomorphisms and scalings of the metric via the DeTurck trick to simplify the equation and compute the principal symbol of the relevant operator. In Sec.IV, we study the linearized stability of the flat space which is a critical point of the flow and find an unstable mode. In Sec.V, we refine the gradient flow formulation of the Cotton flow, and expand the entropy functional for a generic fixed point of the flow up to second order in perturbation theory, and show that Einstein spaces as critical points are unstable. In  Sec VI, we give details of the properties of extended Cotton solitons and show that there are no non-trivial compact solutions. We give examples of Cotton and Ricci solitons as well as Topologically Massive solitons.  In the Appendices A,B,C, we expound upon some of our calculations and in Appendix D, we show how the Cotton tensor flow under arbitarary flows and specifically the Ricci flow.

\section{Linearization of the Cotton Flow about an Arbitrary Curved Background}

Cotton flow on a 3 manifold is defined as \footnote{One is almost tempted to call this "100 \% Cotton flow," as there will be non-pure, mixed flows that we shall discuss.}
\begin{equation}
 \partial_t g_{ij}=  C_{ij},
\label{cotflow1}
 \end{equation}
where we have scaled $t$ to set $\kappa=1$ and from now on we suppress the arguments of the tensors.
The Cotton tensor is given as
\begin{equation}
  C^{ij} = \frac{\epsilon^{ikl}}{\sqrt{g}} \nabla_k \bigg (R^j{_{l}}- \frac{1}{4} \delta^j_l R \bigg ), 
\end{equation}
with $\epsilon^{ikl}$ being the anti-symmetric tensor density defined as $\epsilon^{123}=+1$. [In Appendix A, we compute how some other tensors and scalars built out of tensors flow under Cotton flow.] In what follows we will use the anti-symmetric tensor
\begin{equation}
\eta^{ijk} \equiv \frac{\epsilon^{ijk}}{\sqrt{g}}.
\end{equation}
$C^{ij}$ is symmetric and covariantly conserved $\nabla_i C^{ij}=0$, and traceless $g_{ij}C^{ij}=0$. These constraints tell us that out of 6 components of the (symmetric) metric $g_{ij}$, only 2 can be determined by the Cotton flow equation (\ref{cotflow1}), hence a gauge fixing must be done. To understand the local nature of the Cotton flow about a flat background, equation (\ref{cotflow1}) was linearized
for flat metrics $\bar{g}_{ij}=\delta_{ij}$ in \cite{Cottonflow}. In this section, we will generalize this result for a generic background $\bar{g}_{ij}$ to understand the linearized version of the flow. The results are relevant to the existence and uniqueness of the flow as well as the linear stability of the critical points, namely whether the critical points are saddle points, minima or maxima of the action from which Cotton flow is derived.  For the existence and uniqueness of the flow, unfortunately we have nothing to say since the equation is a third order PDE and necessary mathematical technology does not seem to exist yet, but stability issue will be studied in the next section.

To linearize the Cotton tensor about an arbitrary background $\bar{g}_{ij}$, it is useful to recast it in an explicitly symmetric form as \cite{DeserTekinTMG} 
\begin{equation}
  2 C^{ij}=\eta^{ikl} \nabla_k G^j{_l} +\eta^{jkl} \nabla_k G^i{_l},
\end{equation}
where $ G^j{_l}= R^j{_l}-\frac{1}{2} \delta^j{_l}R$ is the Einstein's tensor. Assuming $g_{ij}=\bar{g}_{ij}+h_{ij}$, and $h_{ij}$ is small compared to the background metric $\bar{g}_{ij}$ which at this stage arbitrary (namely not a critical point of the flow), then one obtains the linearized Cotton tensor as
 \begin{equation}
\begin{aligned}
 2(C^{ij})_L=&-\frac{3h}{2}\,{\bar{C}^{ij}}-\frac{1}{2}\eta^{ikl}\,\bar{\square}\bar{\nabla}_{k}h^{j}{_{l}}
 +\frac{1}{2}\eta^{ikl}\,\bar{\nabla}^{j}\bar{\nabla}_{n}\bar{\nabla}_{k}h_{l}{^{n}}
+ \frac{3}{2}\eta^{ikl}\, \bar{\nabla}_{k}(\bar{{\cal S}}^{nj} h_{nl})
 +\frac{1}{6} \eta^{ikl}\, \bar{R} \bar{\nabla}_k h^j{_l}\\
 &-\frac{1}{2}\eta^{ikl}\,\bar{{\cal S}}^j{_l} \bar{\nabla}_{k}h -\frac{1}{2}\eta^{ikl}\, h^n{_l} \bar{\nabla}_n \bar{{\cal S}}^j{_k}+\eta^{ikl}\,\bar{{\cal S}}_{nk}{\nabla}^j h^n{_l}+\eta^{ikl}\,\bar{{\cal S}}_l{^n}{\nabla}_{n}h^{j}{_{k}}+i \leftrightarrow j,
 \label{cottlin}
 \end{aligned}
 \end{equation}
where $\bar{{\cal S}}_{ij}=\bar{R}_{ij}-\frac{1}{3} \bar{g}_{ij} \bar{R}$ is the traceless Ricci tensor and $\bar{\square}=\bar{g}^{ij}\bar{\nabla}_i\bar{\nabla}_j$ and $h \equiv \bar{g}^{ij} h_{ij}$. 
Note that "\emph{L}" refers to the linearization and all the barred quantities are taken with respect to the background metric, which also raises and lowers the indices of linearized quantities. One must be careful with the linearization of the up and down indices, for example
\begin{equation}
 (C_{ij})_L= (C^{kl})_L\,\bar{g}_{ik}\bar{g}_{jl}+\bar{C}^k{_j}h_{ik}+\bar{C}^k{_i}h_{jk}.
\end{equation}

Derivation of (\ref{cottlin}) is somewhat tedious, hence we relegate the details into Appendix B. This is the main equation that should be used in the study of the stability of various critical points of the flow. Let us note several properties of (\ref{cottlin}). First of all one obtains the contraction $\bar{g}_{ij} (C^{ij})_L=-h_{ij}\bar{C}^{ij}$, as expected from the linearized tracelessness condition $(g_{ij}C^{ij})_L=0$. Secondly, the linearized version of the divergence-free condition of the Cotton tensor reads at this order as
\begin{equation}
 \bar{\nabla}_i (C^{ij})_L+(\Gamma^i{_{ik}})_L\bar{C}^{kj}+(\Gamma^j{_{ik}})_L\bar{C}^{ik}=0,
\end{equation}
which is satisfied by (\ref{cottlin}) which can be shown after a rather long computation.
 
Let us also understand the symmetries of the linearized Cotton tensor (\ref{cottlin}). Under infinitesimal diffeomorphisms and scalings of the metric as   
  \begin{equation}
   \delta_{\zeta,\lambda} h_{ij} =\bar{\nabla}_i \zeta_j+\bar{\nabla}_j \zeta_i+\lambda(x) \bar{g}_{ij},
  \label{difftraa}
  \end{equation}
with $\lambda(x)$ a scalar function and $\zeta_i$ a vector field, (\ref{cottlin}) transforms as
\begin{equation}
 \delta_{\zeta,\lambda} (C^{ij})_L=\zeta_k \bar{\nabla}^k \bar{C}^{ij}-\bar{C}^{k j} \bar{\nabla}_k \zeta^i-\bar{C}^{k i} \bar{\nabla}_k \zeta^j-\frac{5}{2} \lambda(x) \bar{C}^{ij},
\label{difflincota}
 \end{equation}
 where the first three terms come from the Lie derivative ${\cal L}_{\zeta} (C^{ij})_L$ as expected. The important point is that (\ref{difflincota}) vanishes at the fixed points ($\bar{C}^{ij}=0$) of the flow, hence diffeomorphism and scalings apparently make the critical point a \emph{saddle point}. But this is actually a red-herring coming from the symmetry of the theory and so must be moded out or gauged out to study genuine flows. This property of the linearized flow equation can also be reached in the non-linear flow equation. Diffeomorphism are easy to understand (they are similar to the Ricci flow case), therefore let us concentrate on the scalings of the metric.

Consider changing the $t$-parameter to $s \rightarrow s(t)$, and define a new metric on the manifold as 
\begin{equation}
 \tilde{g}_{ij}(s) \equiv \varphi(s) \, g_{ij}(t(s)),
\end{equation}
such that $ g_{ij}(t)$ satisfies the Cotton flow equation
\begin{equation}
\partial_t g_{ij}(t)=C_{ij}(t). 
\end{equation}
Then, lets check the flow satisfied by $\tilde{g}_{ij}(s)$:
\begin{equation}
 \frac{d}{ds} \tilde{g}_{ij}(s)= \tilde{g}_{ij}\frac{1}{\varphi(s)} \frac{d\varphi(s)}{ds}+\varphi(s)\frac{d t}{ds} C_{ij}(t).
\end{equation}
Since the Cotton tensor transforms under conformal scalings as $\tilde{C}_{ij}(\tilde{g})=\varphi^{-\frac{1}{2}}C_{ij}(g)$, (namely $\sqrt{g}\,C^i{_j}$ is conformally invariant) one has the following flow equation
\begin{equation}
\frac{d}{ds} \tilde{g}_{ij}(s)=\frac{d(\log \varphi(s))}{ds} \tilde{g}_{ij} +\varphi^{\frac{3}{2}}(s)\frac{dt}{ds} \tilde{C}_{ij}(\tilde{g}).
\end{equation}
Hence, choosing $\varphi^{3/2} \frac{dt}{ds}=1$ yields
\begin{equation}
\frac{d}{ds} \tilde{g}_{ij}(s)=\frac{d(\log \varphi(s))}{ds} \tilde{g}_{ij} +\tilde{C}_{ij}(\tilde{g}).
\end{equation}
Therefore by parameterizing the time variable one can produce the scaling term.

Next having observed the symmetry of the flow, we can introduce the DeTurck trick \cite{DeTurck} which was used to show that the Ricci flow is parabolic and hence diffusive. Without this trick, parabolic nature of the Ricci flow for an arbitrary initial metric is a highly cumbersome task to show \cite{Hamilton1}. Since the Cotton tensor is of third order, one should not expect to apply the theory of elliptic differential operators. But, as we shall see, a modified version of DeTurck trick removes the zero modes of the relevant operator at the critical point and hence modes out the symmetries of the flow, yielding a somewhat simpler equation.

Before closing this section let us briefly note how the Cotton flow equation should be linearized in the first order formalism where instead of the metric one uses the dreibein and the spin connection.
 For this purpose, let  $e^a$ and $w^a{_b}$ denote the dreibein and spin-connection $1$-form respectively. Then the curvature $2$-form reads 
\begin{equation}
 R^{ab}=dw^a{_b}+w^a{_c}\wedge w^{cb},
\end{equation}
from which the Ricci $1$-form and the scalar curvature can be computed via the inner product. 
\begin{equation}
 (Ric)^a=\iota_b R^{ba}, \hskip .5 cm R=\iota_a\iota_bR^{ba}.
\end{equation}
Then the Cotton $2$-form reads
\begin{equation}
 C^a=DY^a \equiv dY^a+w^a{_b}\wedge Y^b, 
\end{equation}
with the Schouten 1-form given as $Y^a= (Ric)^a-\frac{1}{4}Re^a$. With these definitions, the first order form of the Cotton flow (which was employed in \cite{Cottonflow}) to discuss the flow of homogeneous quantities reads 
\begin{equation}
 \partial_t e^a=\ast C^a,
\end{equation}
where $\ast $ denotes the Hodge dual. Assuming $\bar{e}^a$ satisfies 
\begin{equation}
\partial_t\bar{e}^a=\ast \bar{C}^a=0,
\end{equation}
following the notation of \cite{Cebeci}, let us expand around the critical point of the flow as
\begin{equation}
 e^a\equiv\bar{e}^a+\varphi^a{_b}\bar{e}^b.
\end{equation}
Here  $\bar{\iota}_b\bar{e}^a=\delta^a_b$. The linearized flow equation reads
\begin{equation}
\bar{e}^b (\partial_t \varphi^a{_b})=\bar{\ast} (C^a)_L.
 \end{equation}
To compute the linearized Cotton $2$-form, let us first find the linearized spin connection 
\begin{equation}
w^a{_b}=\bar{w}^a{_b}+\bar{e}^c \Big( \bar{D}_b\varphi^a{_c}-\bar{D}_a\varphi_{bc} \Big),
\end{equation}
where $\bar{D}_a\equiv\bar{\iota}_a \bar{D}$ and the background spin connection satisfying $\bar{D}\bar{e}^a=d\bar{e}^a+\bar{w}^a{_b}\wedge \bar{e}^b=0$. One obtains the curvature $2$-form as
 \begin{equation}
  R_{ab}=\bar{R}_{ab}-\bar{e}^c\wedge\bar{D}\Big (\bar{D}_b\varphi_{ac}-\bar{D}_a\varphi_{bc} \Big),
 \end{equation}
and the Ricci 1-form as 
 \begin{equation}
(Ric)^a = (\bar{Ric})^a+\bar{e}^c \Big(\bar{D}^2 \varphi^a_{c}-\bar{D}_b \bar{D}^a \varphi^b{_c}  \Big )+\bar{D}\Big(\bar{D}^a \varphi -\bar{D}^b \varphi_{ba} \Big )-\varphi_b{^c}\bar{\iota}_c \bar{R}^{ba},
 \end{equation}
where $\bar{D}^2=\bar{D}_a\bar{D}^a$, from which the linearized Schouten and Cotton forms can be computed.

\section{DeTurck Trick in the Cotton flow}
 
By moving the $\bar{\square}$ operator to the left in the second term and playing with the relevant indices, we have deliberately arranged (\ref{cottlin}) in such a way that we can gauge away some terms by extending the Cotton flow as
\begin{equation}
\partial_t g_{ij} =C_{ij}+\nabla_i V_j+\nabla_j V_i.
\label{deturckflw}
\end{equation} 
As in the Ricci flow case we can take the symmetry breaking vector as 
\begin{equation}
 V_i= g_{ik}\,g^{mn}\Big(\Gamma^k_{m n}-\tilde{\Gamma}^k_{m n}\Big),
\end{equation}
but to simplify our equation further by removing the third term and its partner in (\ref{cottlin}) we will take it as 
 \begin{equation}
  V_i \equiv -\frac{1}{4} \eta_i{^k}{_l}\, \nabla^m \Big(\Gamma^l_{k m}-\tilde{\Gamma}^l_{k m}\Big)+ g_{ik}\,g^{mn}\Big(\Gamma^k_{m n}-\tilde{\Gamma}^k_{m n}\Big),
  \label{deturck1}
 \end{equation}
where $ \tilde{\Gamma}^\alpha_{l \lambda}$ is a fixed connection on the manifold which we shall take it to be coming from the initial metric $\bar{g}_{ij}$. There is a subtle issue here which can be seen for the flat ($\bar{R}_{ij}$) critical points: One might mistakingly think that the first piece in (\ref{deturck1}) would be sufficient to remove all the zero modes but it turns out that it only removes the diffeomorphisms, but not the scalings $\delta_{\lambda}h_{ij}= \lambda(x)\delta_{ij}$. Linearization of (\ref{deturck1}) yields
 \begin{equation}
  (V^i)_L= -\frac{1}{4} \eta^{ik}{_l}\, \bar{\nabla}^m \bar{\nabla}_k h_m{^l}+\bar{\nabla}_m h^{im}-\frac{1}{2}\bar{\nabla}^i h.
 \end{equation}
Since one has $\partial_t g^{ij} =-(C^{ij}+\nabla^i V^j+\nabla^j V^i)$, the third term in (\ref{cottlin}) and its symmetric partner are canceled with this gauge-fixing, DeTurck-Cotton flow becomes 

\begin{equation}
  \partial_t h^{ij}=-\frac{1}{4}\eta^{ikl}\,\bar{\square}\bar{\nabla}_{k}h^{j}{_{l}}+i \leftrightarrow j+\mbox{Reaction Terms}
  \label{cotdeturck},
\end{equation}
where the reaction terms do not contribute to the principal symbol of the operator on the left-hand side. Of course, even after this \emph{simplification} one cannot say much about the diffusive nature of the equation, since the principal symbol of the relevant operator vanishes as we show here.

Let us calculate the principal symbol of the Cotton tensor as an operator on the metric using the above linearized form. We have 
\begin{equation}
 \sigma[C](\varphi) (h)_{ij}=-\frac{1}{4} \eta_i{^{kl}}\varphi^2\varphi_kh_{jl}-\frac{1}{4} \eta_j{^{kl}}\varphi^2\varphi_kh_{il},
\end{equation}
with $\varphi^2=g^{ij}\varphi_i\varphi_j$ and $\varphi_i$ is a non-vanishing vector field. Then the inner product vanishes
\begin{equation}
\Big < \sigma[C](\varphi) (h)_{ij}, h_{ij}\Big>=0.
 \label{principalsy}
\end{equation}
This class of the operators are known as sub-elliptic operators when the operator is of second order. The classification for third order operators has not been done to the best of our knowledge, so we have not much to conclude from (\ref{principalsy}).

As we noted, the critical points of the flow satisfy $\bar{C}_{ij}=0$, that is, they are the locally conformally flat metrics. Among these critical points, one clearly has Einstein spaces $\bar{{\cal S}}_{ij}=0$ as a subclass, including the flat space. In 3 dimensions, since the Riemann tensor can be written in terms of the Ricci tensor and scalar curvature, Ricci flat metrics are Riemann flat. In the next section we study the stability of flat metrics as a subclass of critical points of the flow.

\section{STABILITY ANALYSIS OF THE FLAT SPACE}
Let us now study the \emph{stability} of the flat space in Cotton flow. This problem is interesting on its own since flat space is a fixed point of the flow, but it is also relevant to understand the stability of \emph{any} manifold under small wave-length (or high momentum) perturbations. Here what we mean by small wave-length perturbation is that $\frac{\lambda}{\lvert R_{ij}\rvert}\ll1$, with $\lvert R_{ij}\rvert$ a typical curvature of the manifold in some coordinates. Namely for these sufficiently high energy modes, any space is locally flat. One expects that these modes die out, or diffuse as time passes if the critical point is a stable point under linear perturbations. But it will turn out that there is a potentially unstable mode in flat space at the linearized level.

For the flat background, the linear Cotton flow equation with the DeTurck terms becomes \footnote{See Appendix C for another computation of the linearized instability.}
\begin{equation}
 \partial_t h_{ij}=-\frac{1}{4}\eta_i{^{kl}}\,
  \partial^2\partial_{k}h_{j}{_l}
+\partial_i\partial^k h_{kj}-\frac{1}{2}\partial_i\partial_jh+i \leftrightarrow j,
  \label{cottonflatss}
\end{equation}
where $\partial^2$ is the usual Laplacian in flat space. In a broad sense, this equation is a tensor version of the \emph{linearized} Korteweg-de Vries (KdV) equation albeit an anisotropic one, in the sense that the flow of $h_{ij}$ with given $i$ and $j$ is determined by $h_{mn}$ with  $m$ and $n$ not necessarily equal to  $i$ and $j$. This anisotropy is extremely crucial for the diffusive or non-diffusive nature of the equation since the sign of $h_{ij}$ is not necessarily positive. Therefore, in the stability analysis, the sign of the amplitude of the perturbation is also relevant. So to see the nature of this equation, we must first diagonalize it. 

Given the initial perturbation at $ t=0 $ as $h_{ij}(0, {\bf x})$ with the asymptotic fall off as $\lim_{\lvert {\bf x} \rvert \to\infty} h_{ij}(0, {\bf x}) \to 0 $, we can construct ``plane-wave'' type solutions or Fourier modes from which we can construct the general solution that satisfies the asymptotic boundary condition.

Let us insert the plane-wave ansatz into (\ref{cottonflatss})  
\begin{equation}
h_{ij}(t, {\bf x} ) = \xi_{ij}({\bf p})  e^{i({\bf p}.{\bf x}-w({\bf p}) t)},
\label{ttflathgzg}
\end{equation}
where $\xi_{ij}({\bf p}) $ are (complex) polarizations which only depend on the mode (or momentum ${\bf p}$) and $w=w({\bf p})$ is the dispersion relation to be determined below. This insertion leads to a consistency condition on the polarizations
\begin{equation}
\xi_{ij}= -\frac{p^2}{4 w} p_k\eta_i{^{kl}}\,\xi_{jl}-\frac{i}{w}p_ip^k\xi_{kj}+\frac{i}{2w}p_ip_j\xi
+i \leftrightarrow j,
  \label{ttflathggd1}
\end{equation}
where $p= \lvert {\bf p} \rvert $ and the trace of the polarization tensor is $\xi=\delta_{ij}\xi^{ij}$. We will now play with this equation to find $w=w({\bf p})$ and the constraints on the polarization tensor. [Note that these constraints come from the linearized Bianchi identity and the tracelessness of the Cotton tensor.] Taking the trace of (\ref{ttflathggd1}), one obtains
\begin{equation}
\Big(1-\frac{i}{w}p^2\Big)\xi=-\frac{2i}{w}p^ip^j\xi_{ij},
\label{tracemode1}
\end{equation}
and contracting  (\ref{ttflathggd1}) with $p^ip^j$ yields
\begin{equation}
\Big(1+\frac{2i}{w}p^2\Big)p^ip^j\xi_{ij}=\frac{i}{w}p^4\xi.
\label{contractmode}
\end{equation}
To get the possible dispersion relations, let us \emph{assume} that $\xi \ne0$ and $w=ip^2$ then from (\ref{tracemode1}) $p^ip^j\xi_{ij}=0$ is obtained. But (\ref{contractmode}) gives $\xi=0$ so this is a contradiction which can be resolved by assuming $w=ip^2$ and $\xi=0$ and $p^ip^j\xi_{ij}=0$. This necessarily says that $\xi_{ij}=\xi^{TT}_{ij}$, that is $\xi_{ij}$ is a \emph{transverse traceless} tensor. Going back to (\ref{ttflathggd1}), one has
\begin{equation}
\xi^{TT}_{ij}= -\frac{p^2}{4 w} p_k \Big(\eta_i{^{kl}}\,\delta^m{_j}
  +\eta_j{^{kl}}\,\delta^m{_i}\Big)\xi^{TT}_{ml},
  \label{ttflathggd}
\end{equation}
This equation can be diagonalized once the left-hand side is plugged to the $\xi^{TT}_{ml}$ in the right-hand side and contraction of $\eta$'s are carried out, to get\footnote{Alternatively, for $TT$-modes, by taking one more derivative of (\ref{cottonflatss}), one 
obtains
\begin{equation}
 \partial^2_t h_{ij}=-\frac{1}{4}(\partial^2)^3 h_{ij},
\end{equation}
whose Fourier transform yields (\ref{dispertt}).}
 \begin{equation}
 \Big(w^2+\frac{p^6}{4}\Big)\xi^{TT}_{ij}=0.
\end{equation}
Since we cannot take $\xi^{TT}_{ij}$ to be zero (otherwise $h_{ij}\equiv0$), we have
\begin{equation}
w=\pm \frac{i}{2} p^2 \lvert {\bf p} \rvert,
\label{dispertt}
\end{equation}
which contradicts with our earlier assumption $w=ip^2$. Hence we conclude that the assumed dispersion relation is not correct: $w\ne ip^2$.

Clearly the same reasoning works for the other possible dispersion relations $w=-2ip^2$ coming from (\ref{contractmode}). Namely, two apparently possible dispersion relation in (\ref{tracemode1}) and (\ref{contractmode}) are do not yield possible solutions of (\ref{ttflathggd1}). Hence we must assume $w\ne ip^2$ and $w\ne -2ip^2$. Then, assuming $\xi \ne0$ and $p^ip^j\xi_{ij}\ne 0$ equations (\ref{tracemode1}) and (\ref{contractmode}) yield 
\begin{equation}
\Big (1+\frac{ip^2}{w}\Big)\xi=0,
\end{equation}
which leads to another possible dispersion relation 
\begin{equation}
w=-ip^2,
\end{equation}
and the identity $\xi=p^ip^j\xi_{ij}$. Now, let us evaluate whether this mode is viable or not. For this purpose decompose $\xi_{ij}$ as follows
\begin{equation}
\xi_{ij}=\xi^{TT}_{ij}+\Big(p_ip_j-\frac{1}{3}\delta_{ij}p^2\Big)A(p)+\frac{1}{3}\delta_{ij}\xi,
\label{decompose11}
\end{equation}
where $A( p)$ is arbitrary at this stage. Using $\xi=p^ip^j\xi_{ij}$ in (\ref{decompose11}) gives $\xi=A(p)p^2$, and so one arrives at
\begin{equation}
\xi_{ij}=\xi^{TT}_{ij}+\frac{p_ip_j}{p^2} \xi.
\label{concludedecomp}
\end{equation}
 Substituting (\ref{concludedecomp}) into (\ref{ttflathggd1}) yields
 \begin{equation}
\xi^{TT}_{ij}+\frac{p_ip_j}{p^2}\xi= -\frac{p^2}{4 w} p_k\eta_i{^{kl}}\,\xi^{TT}_{jl}-\frac{p^2}{4 w} p_k\eta_j{^{kl}}\,\xi^{TT}_{il}-\frac{i}{w}p_ip_j\xi.
\end{equation}  
Inserting $w=-ip^2$, leads to the conclusion that for these modes, the $TT$ polarizations vanish: $\xi^{TT}_{ij}=0$. Therefore this is a valid dispersion relation for the following modes
\begin{equation}
\xi_{ij}=\frac{p_ip_j}{p^2}\xi.
\label{concludedecomp2}
\end{equation}
For these modes $h_{ij}(t,{\bf x})= \xi_{ij}\, e^{i{\bf p}.{\bf x}- p^2 t}$, hence as time passes, these modes decay namely they are diffusive modes.

Let us now consider the final possibility, that is $\xi^{TT}_{ij}\ne0$, then $\xi=0$, $p^ip^j\xi_{ij}=0$, hence we have the complex cubic dispersion relation
\begin{equation}
w=\pm \frac{i}{2} p^2 \lvert {\bf p} \lvert.
\label{dispertt1}
\end{equation}
With this dispersion relation, we still need to make sure that there is a non-trivial solution of equation (\ref{ttflathggd}). We can recast that equation for both signs of the dispersion relation as a matrix equation 
\begin{equation}
\bf \boldsymbol\xi =B \boldsymbol\xi+\boldsymbol\xi B,
\label{mainmatreq}
\end{equation}
with $\boldsymbol\xi=(\xi^{TT}_{ij})$ as a $3 \times 3$ symmetric matrix and $\bf B$ as a $3 \times 3$ anti-symmetric matrix whose elements are given as
\begin{equation}
B_{ij} \equiv  \frac{p^2 }{4 w } p_k \eta_{ij}{^k}.
\end{equation}
Equation (\ref{mainmatreq}) is known as Sylvester's equation which can be solved \cite{MatrixEqn} but we do not need the general solution, all we need is a non-trivial solution which we can find as follows. Suppose ${\bf p}=(0,0,p)$, then, $\boldsymbol\xi$ and the $\bf B$ matrix read as
\[ \boldsymbol\xi= \left( \begin{array}{ccc}
\xi_{11} & \xi_{12} & 0 \\
\xi_{12} & -\xi_{11} & 0 \\
0 & 0 & 0 \end{array} \right), \,\,\,\,\,\,\,\, {\bf B}= \left( \begin{array}{ccc}
0 & p & 0 \\
-p & 0 & 0 \\
0 & 0 & 0 \end{array} \right). \]
 Equation (\ref{mainmatreq}) yields
\begin{equation}
\xi_{12}=\mp i \xi_{11}. 
\end{equation}
Normalizing $\xi_{11}=1, \,\,\, \xi_{12}=\mp i $ depending on the sign of the dispersion $ w=\pm \frac{i}{2} p^2 \lvert {\bf p} \rvert$. Therefore we have non-trivial solutions for both signs of the dispersion relation.

Since the equation (\ref{cottonflatss}) is linear, then the most general solution reads as 
\begin{equation}
h_{ij}(t,{\bf x})= \int d^3 p\,\, \xi_{ij}({\bf p})\, e^{i{\bf p}.{\bf x}\,\pm \frac{1}{2} p^2 \lvert {\bf p} \rvert \,t}.
\end{equation}
We can recast this in terms of the initial perturbation as
\begin{equation}
h_{ij}(t,{\bf x})=\int d^3 x^{'} G({\bf x},{\bf x}^{'},t) \, h_{ij}(0,{\bf x}^{'}), 
\end{equation}
where
\begin{equation}
G({\bf x},{\bf x}^{'},t)= \frac{1}{(2 \pi)^3}\int d^3 p\, e^{i{\bf p}.({\bf x}-{\bf x}^{'})\,\pm \frac{1}{2} p^2 \lvert {\bf p} \rvert \,t}.
\end{equation}
This integral can be evaluated for the \emph{minus} sign, in terms of hyper-geometric and Bessel-kind functions. On the other hand, the \emph{positive} sign mode is problematic. It is clear that the negative sign modes decay in time and hence these perturbations are diffused. As for the positive sign modes, there is a linear instability. But bear in mind that this analysis has been at the perturbative level and the perturbation theory breaks down for these growing modes hence a strict conclusion cannot be made with regard to the non-linear instability of flat space in Cotton flow. 

In fact, even though we have not been able to prove stability at a non-linear level, we conjecture that these growing modes could be tamed at the non-linear level. If this turns out not to be the case, then, since the above analysis is also valid for any manifold for high momentum fluctuations that see the manifold locally flat, any critical point of the flow would be unstable, leading to the conclusion that the theory has no minima but just saddle points. This would be highly unexpected.

\section{Cotton Flow as a Gradient Flow and the Entropy Functional}

In \cite{Cottonflow} a gradient flow formulation of Cotton flow was given which we refine it here to better understand the role of symmetries and the Cotton solitons and also study the issue of stability using the entropy functional. The Chern-Simons action \cite{DJT} on a manifold without a boundary
\begin{equation}
{\cal F}= -\frac{1}{2} \int_{{\cal M}} d^3 x \sqrt{g} \, \eta^{ijk} \Gamma^l_{im} \Big (\partial_j \Gamma^m_{kl}+\frac{2}{3} \Gamma^m_{jn} \Gamma^n_{kl} \Big ),
\label{chssde}
\end{equation}
has the first variation under an \emph{arbitrary} change of the metric as
\begin{equation}
\delta{\cal F}= \int_{{\cal M}} d^3 x \, \sqrt{g} \, C^{ij}\, \delta g_{ij}. 
\label{chseq}
\end{equation}
Observe that for $\delta g_{ij} =\nabla_i V_j+\nabla_j V_i + \lambda(x) g_{ij}$, the first variation vanishes, $\delta{\cal F}=0$, owing to the fact that Cotton tensor is traceless and covariantly conserved. Hence such a variation gives us the symmetries of the theory:
which are diffeomorphisms and arbitrary scalings of the metric. To get a steepest descend, we can choose the following flow
\begin{equation}
 \delta g_{ij}\equiv\partial_tg_{ij} =C_{ij}+\nabla_i V_j+\nabla_j V_i + \lambda(x) g_{ij},
 \label{newcotkglthdfgy}
\end{equation}
which leads to
\begin{equation}
\delta{\cal F} \equiv \frac{d{\cal F}}{d t}= \int_{{\cal M}} d^3 x \, \sqrt{g} \, C_{ij}C^{ij}.
\label{flowctson}
\end{equation}
To get a gradient flow we must require $ \frac{d{\cal F}}{d t} \ge 0 $ where the bound is saturated for the critical points that satisfy 
the following equation
\begin{equation}
C_{ij}+\nabla_i V_j+\nabla_j V_i + \lambda(x) g_{ij}=0,
\label{cotsolson}
\end{equation}
which we shall call it as the {\emph{extended Cotton solitons}}. [In \cite{Cottonflow}, the case of constant $\lambda$ metrics were called Cotton solitons]. It is clear that (\ref{flowctson}) vanishes at the critical points, since by definition $\delta g_{ij}=0$ there. Requiring the volume density to be a conserved quantity of the flow $\partial_t \sqrt{g}=0$, yields, from (\ref{newcotkglthdfgy}), the constraint
\begin{equation}
2 \nabla_i V^i +3 \lambda(x)=0,
\end{equation}
for all metrics $g_{ij}$ that are solutions of the flow.

To understand the stability of the fixed points, let us expand the Entropy functional of the Cotton flow up to the second order in $\delta g_{ij}$ for a fixed time $t$. Then from (\ref{chseq}), one obtains
\begin{equation}
 \delta^2  {\cal F}= \int_{{\cal M}} d^3 x \sqrt{\bar{g}} \, \Big(\frac{1}{2} \bar{g}^{ij} \delta g_{ij} \bar{C}^{kl} \delta g_{kl}+ (C^{ij})_L \delta g_{ij} \Big),
\end{equation}
where $\bar{g}_{ij}$ refers to the critical points that satisfy (\ref{cotsolson}). Making use of (\ref{cottlin}) one arrives at 
\begin{equation}
\begin{aligned}
 \delta^2  {\cal F}=\int_{{\cal M}} d^3 x \sqrt{\bar{g}}\bigg (&-\frac{h}{2} h_{ij}\bar{C}^{ij}-\frac{1}{2} h_{ij}h^j{_n} \bar{C}^{i n}
 +\frac{1}{2} \eta^{ikl}\,h_{ij} \bar{\nabla}_{k} \triangle^{(2)}_L h^{j}{_l}+\frac{1}{2}\eta^{ikl}\, h_{ij}\bar{\nabla}^{j}\bar{\nabla}_k \bar{\nabla}_n h_{l}{^{n}}\\
&-\frac{1}{2} \eta^{ikl}\,h \bar{{\cal S}}_{l}{^j} \bar{\nabla}_k h_{ij} 
-\frac{1}{2} \eta^{ikl}\,h^{j n} \bar{{\cal S}}_{k n} \bar{\nabla}_l h_{i j}+\eta^{ikl}\,h_{ij} \bar{{\cal S}}_{k}{^j} \bar{\nabla}^n h_{n l}\\
&+\frac{1}{2}\eta^{ikl}\,h_{ij} \bar{{\cal S}}_{n l} \bar{\nabla}^n h^j{_k} +\frac{1}{2}\eta^{ikl}\,h_{ij} \,\bar{{\cal S}}_{k n} \bar{\nabla}^j h^n{_l}
+\frac{1}{3}\eta^{ikl}\,h_{ij} \bar{R} \bar{\nabla}_l h^j{_k} \bigg),
\label{stabint}
\end{aligned}
\end{equation}
where we have dropped some boundary terms and made use of the Lichnerowicz operator action on a symmetric two tensor as
\begin{equation}
 \triangle^{(2)}_L h_{ij}=-\bar{\square} h_{ij}-2\bar{R}_{ikjl}h^{kl}+2 \bar{R}^k{_{(i}}h_{j) k},
\end{equation}
which yields in three dimensions
\begin{equation}
  \triangle^{(2)}_L h_{ij}= -\bar{\square} h_{ij}-2 \bar{g}_{ij}\bar{R}_{kl} h^{kl}-2 h \bar{R}_{ij}+\bar{g}_{ij} h \bar{R}- \bar{R} h_{ij} +3 \bar{R}_{ik}h_j{^k}+3 \bar{R}_{jk}h_i{^k}.
\label{lich}
\end{equation}
To ensure the stability of the critical points one must have 
\begin{equation}
 \delta^2  {\cal F}\biggr\rvert_{\bar{g}_{ij}} > 0.
\end{equation}
For a generic conformally flat fixed point, or for Cotton soliton backgrounds, even though it is possible that $\delta^2 {\cal F}$ is positive, we have not been able to show this. Instability of Einstein spaces $\bar{{\cal S}}_{ij}=0$, can be seen quite easily if one decomposes the perturbation as
\begin{equation}
 h_{ij}=h^{TT}_{ij}+\bar{\nabla}_i\varphi_j+\bar{\nabla}_j\varphi_i+f(x)\bar{g}_{ij},
\end{equation}
which after integration by parts reduces the action (\ref{stabint}) to
\begin{equation}
 \delta^2  {\cal F}=\frac{1}{2}\int_{{\cal M}} d^3 x \sqrt{\bar{g}}\,\eta^{ikl}\,\bar{\nabla}_{k}h^{TT}_{ij} \Big(\bar{\square}-\frac{\bar{R}}{3}\Big) h^{TT j}{_l},
\end{equation}
where $\bar{R}$ is constant. The first derivative on the perturbation leads to a linear instability. This is consistent with  our earlier flat space analysis. Hence Einstein spaces, including the flat space, critical points of the flow are linearly unstable.
This is almost evident but let us see it more explicitly for the $S^3$ metric in the following coordinates
\begin{equation}
ds^2=\cos^2 \rho \, d \tau^2 +\sin^2 \rho \, d \phi^2+ d \rho^2, 
\end{equation}
with $\bar{R}=6$. Assuming
\[ h^{TT}_{ij}= \left( \begin{array}{ccc}
h_{11} & h_{12} & 0 \\
h_{12} & -h_{11} & 0 \\
0 & 0 & 0 \end{array} \right), \]
second variation of the entropy about $S^3$ for this perturbation yields 
\begin{equation}
\delta^2 {\cal F}=-2\int_{{\cal M}} d^3 x \, h_{11}\Big (\bar{\square}-3 \Big) \, \partial_3 h_{12},
\end{equation}
which is not positive in general. Thus, $S^3$ is a saddle point not a minimum of the flow. In \cite{Cottonflow}, it was shown that as a critical point  $S^3$ is stable under homogeneous deformations.

\section{Extended COTTON SOLITONS}

In the previous section, we have shown that generic fixed points of Cotton flow are {\emph {extended Cotton solitons}} that satisfy
\begin{equation}
C_{ij}+\nabla_i V_j+\nabla_j V_i + \lambda({ x}) g_{ij}=0,
\label{cotsolsonsedr}
\end{equation}
where $\lambda({x})$ is not necessarily constant as was assumed in \cite{Cottonflow}. As was shown in \cite{Cottonsoliton1}, if 
$\lambda({x})$ is constant and the manifold is compact without a boundary, then there are no non-trivial Cotton soliton solutions, namely $\lambda=0$, $C_{ij}=0$, and $V_i$ is a Killing vector
\begin{equation}
{\cal{L}}_V g=0.
\end{equation} 
For extended Cotton solitons, let us prove a similar theorem.
\newline
\newline
{\bf Theorem:} \emph{On a compact Riemannian manifold without boundary, all solutions of (\ref{cotsolsonsedr}) have ${\cal{L}}_V g=-\lambda g$, namely, $V$ is a conformal Killing vector and $C_{ij}=0$: There are no non-trivial extended Cotton solitons.}
\newline
\newline
{\bf Proof:} The square of the Lie derivative of the metric reads
\begin{equation}
\lvert{\cal{L}}_V g\rvert^2=\square (V_i V^i)-2V_i\square V^i+2\nabla_i (V^j\nabla_j V^i)-2V_i\nabla^i\nabla_jV^j-2V^iV^j
R_{ij}.
\label{difcot}
\end{equation}
Trace and divergence of (\ref{cotsolsonsedr}) can be computed as
\begin{equation}
2\nabla_iV^i+3\lambda (x)=0,
\label{trcotsol}
\end{equation}
\begin{equation}
\square V_i+R^j{_i}V_j-\frac{1}{2}\nabla_i\lambda (x)=0.
\label{divcotsol}
\end{equation}
Making use of these in (\ref{difcot}), one arrives at
\begin{equation}
\lvert{\cal{L}}_V g\rvert^2=\square (V_i V^i)+2\nabla_i(V^i\lambda)+2\nabla_i (V^j\nabla_j V^i)+3\lambda^2(x).
\end{equation}
Integrating this over the manifold, once the boundary terms are dropped, yields
\begin{equation}
\int d^3x\sqrt{g}\,\lvert{\cal{L}}_V g\rvert^2=3\int d^3x \sqrt{g}\,\lambda^2(x),
\end{equation}
which can be written as
\begin{equation}
\int d^3x\sqrt{g}\,\lvert{\cal{L}}_V g+\lambda(x) g\rvert^2=0,
\label{desiredextcoty}
\end{equation}
where we made use of (\ref{trcotsol}). Since we are in a Riemannian manifold with a positive metric, from (\ref{desiredextcoty}) it follows that $V_i$ is a conformal Killing vector with the conformal factor $\lambda(x)$. Then from (\ref{cotsolsonsedr}) it follows that $C_{ij}=0$. So on a compact Riemannian manifold without a boundary, there are no non-trivial Cotton solitons besides the conformally flat ones. This proves the theorem. Of course for the Lorentzian signature and the non-compact manifolds, the theorem does not follow.

Let us study some properties of extended Cotton solitons. Taking the divergence of (\ref{divcotsol}) and making use of (\ref{trcotsol}) one arrives at
\begin{equation}
\nabla_i \Big(R^i{_j}V^j-\nabla^i\lambda(x)\Big)=0,
\label{conservedcurrent}
\end{equation}
which can be recast as 
\begin{equation}
\partial_i \Big ( \sqrt{g}(R^i{_j}V^j-\partial^i\lambda(x)) \Big)=0,
\label{partialconserved}
\end{equation}
hence $J^i\equiv \sqrt{g}(R^i{_j}V^j-\partial^i\lambda)$ is a "conserved current". For some manifolds one can define the following conserved \emph{total} charge. Suppose two space $\Sigma$ foliate the three manifold ${\cal M}$ and let $\hat{n}^i$ be the normal to the surface $\Sigma$. And let $\gamma_{ij}$ denotes the induced metric on the $\Sigma$. Then
\begin{equation}
Q\equiv \int_\Sigma d^2x \sqrt{\gamma}\, \hat{n}_i \, \Big (R^i{_j}V^j-\partial^i\lambda(x) \Big), 
\end{equation}
is a conserved total charge of the manifold, if the following condition is satisfied. Say we separate one of the coordinates as "$r$" and $\Sigma$ as the constant $r$ surface. Then for large $r$, one has
\begin{equation}
J^i= R^i{_j}V^j-\partial^i\lambda(x) \underset{\mbox{large}\, \,\,r}{
 \xrightarrow{\hspace*{1.1cm}} }\frac{1}{r^{1+\epsilon}}\,, \hskip 1cm \epsilon>0,
\end{equation}
for $i \ne r$. This leads to
\begin{equation}
 \frac{dQ}{dr}=0.
\end{equation}
While, at the moment, we do know explicit examples of extended Cotton solitons, (one could perhaps find these using the method of \cite{Gurses:2008wu} since these are less restrictive solutions compared to the Cotton solitons with constant $\lambda$). In fact let us give an example in the Riemannian setting with a constant $\lambda$ \cite{Cottonsoliton1}. For constant $\lambda$,  there are many in the Lorentzian setting \cite{Cottonsoliton1, Cottonsoliton2, Cottonsoliton33}.

Consider the following metric
\begin{equation}
 ds^2=dx^2+dy^2+\Big(\frac{1}{\ell}(y dx-x dy)+dz\Big)^2,
 \label{heisenberg}
\end{equation}
with $\ell >0$. This metric has the following properties
\begin{equation}
\begin{aligned}
 &i)\,\,\,g=\mbox{det}[g_{ij}]=1, \hskip 1cm ii)\,R=-\frac{2}{\ell^2},\\
 &iii)\,R_{ij}R^{ij}=\frac{12}{\ell^4},\hskip 1.36cm iv)\,C_{ij}C^{ij}=\frac{96}{\ell^6}.
\end{aligned}
\end{equation}
Also, the components of the Cotton tensor can be computed as
\begin{equation}
\begin{aligned}
 C_{11}&=\frac{4}{\ell^3}-\frac{8}{\ell^5}y^2,\hskip .3 cm C_{12}=\frac{8}{\ell^5}xy,\hskip .3 cm C_{13}=-\frac{8}{\ell^4}y,\\
C_{22}&=\frac{4}{\ell^3}-\frac{8}{\ell^5}x^2, \hskip .4 cm C_{23}=\frac{8}{\ell^4}x,\hskip .5 cm C_{33}=-\frac{8}{\ell^3}.
\end{aligned}
\end{equation}

This metric is a Cotton soliton for the scale factor $\lambda=-\frac{16}{\ell^3}$  with the components of the vector field given as
\begin{equation}
(V^i)=\Big(\frac{6}{\ell^3}x-c_1 y+c_3 l,\,\frac{6}{\ell^3}y+c_1 x+c_2,\,\frac{12}{\ell^3}z-\frac{c_2}{\ell} x+c_3 y+c_4 \Big),
\end{equation}
where $c_i$'s are constants. For this metric, one can check that
\begin{equation}
\partial_i(R^i{_j}V^j)=0,
\end{equation}
which is the $\lambda=\mbox{constant}$ version of (\ref{partialconserved}).

It is quite interesting that  (\ref{heisenberg}) is also a Ricci soliton ($R_{ij}+\nabla_i X_j+\nabla_j X_i+\Lambda g_{ij}=0$ ) with the scale factor $\Lambda=\frac{6}{\ell^2}$ and the components of the vector field 
\begin{equation}
(X^i)=\Big(-\frac{2}{\ell^2}x-c_1 y+c_3 l,\,-\frac{2}{\ell^2}y+c_1 x+c_2,\,-\frac{4}{\ell^2}z-\frac{c_2}{\ell} x+c_3 y+c_4 \Big).
\end{equation}
Using this fact we can define a new object that we shall call {\it Topologically Massive solitons} that solve
\begin{equation}
R_{ij}+\frac{1}{\mu} C_{ij}+ \nabla_i W_j+\nabla_j W_i+ P g_{ij}=0,
\end{equation}
which is satisfied by  (\ref{heisenberg}) with the following choices
\begin{equation}
P=\Lambda+\frac{1}{\mu} \lambda, \hskip 1 cm W_i= X_i+\frac{1}{\mu} V_i .
\end{equation}
Here $\mu$ is  the topological  mass parameter.

\subsection*{$PP$-Wave Solution of Cotton Flow}
In the Lorentzian setting, we can show that the $pp$-wave metric, (see \cite{Sisman} for its properties),  
\begin{equation}
ds^2=2du dv+dz^2+2 U(u,z) du^2,
\label{ppwavemetric}
\end{equation}
where $u$ and $v$ are null coordinates. The components of the metric can be written as $g_{\mu \nu}=\eta_{\mu \nu}+2U \lambda_\mu \lambda_\nu$ with the $\lambda_\mu$ vector satisfying the following properties:
\begin{equation}
\begin{aligned}
\lambda^{\mu}\lambda_\mu=0,\,\,\nabla_{\mu}\lambda_\nu=0,\,\,
\,\,\lambda^\mu\partial_\mu V=0,\,\,\lambda^\mu\nabla_\nu\partial_\mu U=0.
\label{plane1}
\end{aligned}
\end{equation}
With these definitions, the Ricci tensor and the Cotton tensor can be computed as
\begin{equation}
R_{\mu\nu}=-\lambda_{\mu}\lambda_{\nu}\partial^2U, \hskip 1 cm C_{\mu\nu}=\epsilon^{\alpha\beta}{_({_{\mu}}\lambda_{\nu)}} \lambda_\alpha\nabla_\beta\partial^2 U.
\end{equation}
The $ pp$-wave metric (\ref{ppwavemetric}) satisfies the Cotton soliton equation with 
\begin{equation}
\begin{aligned}
 &\lambda=0, \hskip .5 cm(V^i)=\Big(0,\,c_1 u+c_3z+c_2, \,-c_3 u-c_4 \Big),\\
 &U(u,z)=\frac{c_1z}{c_3 u+c_4}+\frac{e^{z\sqrt{2(c_3 u+c_4)}}}{\sqrt{2(c_3 u+c_4)}}\,f_1(u)+\frac{e^{-z\sqrt{2(c_3u+c_4)}}}{\sqrt{2(c_3u+c_4)}}\,f_2(u)+f_3(u),
 \end{aligned}
\end{equation}
where $c_i$'s are arbitrary constants and $f_i(u)$'s are arbitrary functions. Interestingly, this Cotton soliton is also a gradient Cotton soliton \cite{Cottonsoliton1} namely $C_{ij}+\nabla_i\nabla_j\psi=0$
\begin{equation}
\psi=\phi(u)+b^2 z, \hskip 1 cm U(u,z)=\frac{e^{z b}}{b}g_1(u)-\frac{e^{-zb }}{b}g_2(u)+g_3(u).
\end{equation} 
One can also show that the pp-wave metric is also a Ricci soliton with the following vector field $X_i$ and the metric function $U$ given as 
\begin{equation}
 (X^i)=\Big(0,\,c_1 u+c_3z+c_2,\,-c_3 u-c_4 \Big), \hskip .3 cm U(u,z)=\frac{c_1 z}{c_3 u+c_4}+\frac{e^{-2z(c_3u+c_4)}}{2(c_3u+c_4)}\,h_1(u)+h_2(u).
\end{equation}
Finally, one can show that (\ref{ppwavemetric}) is a  gradient Ricci soliton with
\begin{equation}
\begin{aligned}
 \psi=\varphi(u)+b z,  \hskip .5 cm U(u,z)=\frac{e^{bz}}{b}f(u)+g(u).
 \end{aligned}
\end{equation}

\section{Conclusions}
We have studied the linearized Cotton flow equation about a generic background with the help of a modified form of the DeTurck trick that removes the zero modes of the relevant operator in the flow equation. Then we specifically studied the perturbations about the critical points of the flow. We found that flat space and Einstein spaces, as critical points of the flow, have unstable modes at the perturbative level, making these spaces saddle points rather than minima. We have also shown that (in Appendix C)  certain conformally flat fixed points that are not Einstein metrics, also have unstable modes with a dispersion relation cubic in the Fourier momentum. We have also supported our arguments by computing the second variation of the entropy functional about the critical points and show that the second variation is negative for Einstein metrics making them saddle points. Finding the true minima (if there is any)  of the Cotton flow equations is an open problem.

We have refined the gradient flow formulation of  Cotton flow and gave a definition of extended Cotton solitons and worked out  some properties of these critical points. We have also given an example of Topologically Massive solitons that is constructed from a solution to both  Ricci and Cotton soliton equations. Our work could be relevant to both mathematics, as discussed in the Introduction and physics.  For example it would be interesting to see the connection between the Cotton flow and various holographic theories that have the pure Chern-Simons action in the bulk of a three manifold \cite{grumiller}.  

\section{Appendix A: Evolution of some quantities tensor under the cotton flow}

Under the Cotton flow, let us note  how some tensors and scalars evolve, 
\begin{enumerate}
 \item The Cotton tensor itself evolves as
\begin{equation}
\begin{aligned}
2 \partial_t C^{ij} = \eta^{mki}&\bigg( \frac{3}{2}\nabla_m(R^j{_ q}C_k{^{q}})+\frac{1}{2}\square \nabla_kC^j{_m}
-\frac{3}{2}R_{k n}\nabla^jC^n{_m} 
+\frac{1}{2}C^n{_m}\nabla_n R^j{_k}\\
&-\frac{1}{2}C^n{_m}\nabla^j R_{nk}
+R^q{_k }\nabla_q C^j{_m}-\frac{1}{2}C^j{_k }\nabla_m R\bigg )+i \leftrightarrow j.
\label{cotevol}
 \end{aligned}
\end{equation} 
\item Square of the Cotton tensor evolves as
\begin{equation}
\begin{aligned}
\partial_t(C_{ij}C^{ij})&=\eta^{mki}\bigg (5C_{ij}C_k{^q}\nabla^jR_{mq}+6C_{ij}R^j{_q}\nabla_mC_k{^{q}}+C_{ij}\square \nabla_kC^j{_m}
-3C_{ij}R_{k n}\nabla_m C^{nj}\\ 
&\hskip 1.3 cm+2C_{ij}R^q{_k }\nabla_q C^j{_m}\bigg)+5C^{ij}C_{im}C^m{_j}.
\end{aligned}
\end{equation} 
\item The Riemann tensor evolves as
\begin{equation}
\begin{aligned}
\partial_t R_{ijkl}&=\frac{1}{2}\epsilon_{jin}\epsilon^{nab}\epsilon_m{^{kl}}\epsilon^{m\sigma\tau}\nabla_\sigma \nabla_a C_{b\tau}
+\frac{1}{2}(R_{kli}{^\rho} C_{j\rho}-R_{klj}{^\rho} C_{i\rho}).
\end{aligned}
\end{equation}
 \item Square of the Riemann tensor evolves as
 \begin{equation}
  \partial_t (R_{ijkl}R^{ijkl})=-4R^{ij}\square C_{ij}-10 R R_{ij}C^{ij}+16 R_{jl} R^{jk}C^l{_k}.
 \end{equation}
 \item Time independence of Bianchi identity $\partial_t(\nabla_iC^{ij})=0$ leads to 
 \begin{equation}
\nabla_iT^{ij}=0, \hskip 1 cm T^{ij} \equiv \partial_t C^{ij}+C^{ik}C_k{^j}-\frac{g^{ij}}{4}C_{kl} C^{kl}.
\end{equation}
 \end{enumerate}
 Note that at the critical points, the flow stops for all the above quantities.
 
\section{Appendix B: LINEARIZATION OF THE COTTON TENSOR AROUND A GENERIC BACKGROUND}
Let us give some details of the calculations leading to equation (\ref{cottlin}). By perturbing the metric about an arbitrary background as $g_{ij}\equiv\bar{g}_{ij}+h_{ij}$, one obtains 
\begin{equation}
 2( C^{ij})_L=-\frac{h}{2}\,{\bar{C}^{ij}}+\underset{A}{\underbrace{\eta^{ikl}\,\bar{\nabla}_{k} (G^j{_l})_L}}+\underset{B}{\underbrace{\eta^{ikl}\,(\Gamma{^{j}}{_{k n}})_L\bar{G}^{n}{_{l}}}}
 +i \leftrightarrow j+{\emph{O}}(h^2).
 \label{Linearcotton}
\end{equation}
From now on we will drop the ${\emph{O}}(h^2)$ terms. The background metric is compatible with the background covariant derivative $\bar{\nabla}_k \bar{g}_{ij}=0$, hence the linearized Christoffel connection reads
\begin{equation}
(\Gamma{^{j}}{_{kn}})_L=\frac{1}{2} \bar{g}^{j m} \Big(\bar{\nabla}_k h_{m n}+\bar{\nabla}_n h_{m k}-\bar{\nabla}_m h_{k n}\Big). 
\label{linchrissybl}
\end{equation}
Our task is to rewrite (\ref{Linearcotton}) in such a way that one can see the terms that can be related to diffeomorphisms of the background metric.
For this purpose, let us first evaluate the $A$ term in  (\ref{Linearcotton}).  Note that the linearization of the Einstein tensor yields
\begin{equation}
\begin{aligned}
 (G^j{_l})_L&=(g^{jm}\,G_{ml})_L
 =-h^{jm}\bar{G}_{ml}+\bar{g}^{jm}\,(G_{ml})_L.
 \label{einsparta}
  \end{aligned}
\end{equation}
Where $(G_{ij})_L=(R_{ij})_L-\frac{1}{2}\bar{g}_{ij}R_L-\frac{1}{2}h_{ij}\bar{R}$. By using the explicit forms of the Linearized Ricci tensor and Ricci scalar
\begin{equation}
\begin{aligned}
 (R_{ij})_L&\equiv \frac{1}{2}\Big(\bar{\nabla}^{k}\bar{\nabla}_{i}h_{j k}+\bar{\nabla}^{k}\bar{\nabla}_{j}h_{i k}
 -\bar{\square}h_{ij}-\bar{\nabla}_{i}\bar{\nabla}_{j}h\Big), \\
 R_L&\equiv(g^{ij} R_{ij})_L=-h^{ij}\bar{R}_{ij}-\bar{\square}h+\bar{\nabla}^{j}\bar{\nabla}^{i}h_{i j},
\end{aligned}
 \end{equation}
one can write the $A$ part as  
\begin{equation}
\begin{aligned}
 \eta^{ikl}\,\bar{\nabla}_{k}(G^j{_l})_L= \eta^{ikl}\,&\bigg(-\bar{\nabla}_{k}(h^{jm}\bar{G}_{ml})+\frac{1}{2}
 \Big(\underset{C}{\underbrace{\bar{\nabla}_{k}\bar{\nabla}_{n} \bar{\nabla}_{l}h^{n j}}}+
 \underset{D}{\underbrace{\bar{\nabla}_{k}\bar{\nabla}_{n}\bar{\nabla}^{j}h_l{^{n}}}}
 -\underset{E}{\underbrace{\bar{\nabla}_{k}\bar{\square}h^j{_l}}}-\bar{\nabla}_{k}\bar{\nabla}_{l}\bar{\nabla}^{j}h \Big)\\
 &-\frac{1}{2}\bar{\nabla}_{k}(h^j{_{l}}\bar{R})\bigg).
\label{Linear}
\end{aligned}
 \end{equation}
Let us now calculate each term of (\ref{Linear}) separately: the $C$ term can be written as
\begin{equation}
\begin{aligned}
  \eta^{ikl}\,\bar{\nabla}_{k}\bar{\nabla}_{n}\bar{\nabla}_{l}h^{n j}=\eta^{ikl}\,\bigg (\bar{R}^j{_{k}}\bar{\nabla}_{n}h_{l}{^{n}}+2\bar{\nabla}_{k}\Big(\bar{R}_{ln}h^{jn} \Big)
  +\bar{\nabla}_{k}\Big(-h\bar{R}_{l}{^{j}}+\bar{R}^{j}{_{n}}h^{n}{_{l}}-\frac{\bar{R}}{2}h^{j}{_l}\Big)\bigg ),
  \label{1}
  \end{aligned}
\end{equation}
where we have used the three dimensional identity 
\begin{equation}
\begin{aligned}
 \bar{R}_{kpqi}=\bar{g}_{kq}\bar{R}_{ip}-\bar{g}_{ki}\bar{R}_{qp}+\bar{g}_{pi}\bar{R}_{qk}-\bar{g}_{pq}\bar{R}_{ik}
-\frac{\bar{R}}{2}\bar{g}_{kq}\bar{g}_{ip}+\frac{\bar{R}}{2} \bar{g}_{ki}\bar{g}_{qp}.
\end{aligned}
\end{equation}
Similarly the $D$ and $E$ terms can be written as
\begin{equation}
\begin{aligned}
 \eta^{ikl}\,\bar{\nabla}_{k}\bar{\nabla}_{n}\bar{\nabla}^{j}h_{l}{^{n}}
 &= \eta^{ikl}\,\bigg (\bar{\nabla}^{j}\bar{\nabla}_n \bar{\nabla}_{k}h_{l}{^{n}}
 -\bar{\nabla}^{j}\Big(\bar{R}^{n}{_{k}}h_{n l} \Big)-\bar{R}^{j}{_{l}}\bar{\nabla}_{n}h_{k}{^{n}} \\
& \hskip 1 cm+\bar{\nabla}_{k}\Big(2h_{n l}\bar{R}^{n j}-h\bar{R}_{l}{^{j}}+h^{j n}\bar{R}_{l n}-\frac{\bar{R}}{2}h^{j}{_{l}}\Big) \bigg ),
\label{2}
\end{aligned}
\end{equation}
\begin{equation}
\begin{aligned}
\eta^{ikl}\,\bar{\nabla}_{k}\bar{\square}h^j{_{l}}&=\eta^{ikl}\,\bigg (\bar{\square}\bar{\nabla}_{k}h^j{_{l}}+2\Big(\bar{R}^j{_{k}}\bar{\nabla}_{n}h^n{_{l}}-\bar{R}_{nk}\bar{\nabla}^{j}h^n{_{l}}\Big)
+\Big(h^n{_{l}}\bar{\nabla}_{n}\bar{R}^j{_{k}}-h^n{_{l}}\bar{\nabla}^{j}\bar{R}_{nk}\Big)\\
&\hskip 1 cm-\bar{R}_l{^{n}}\bar{\nabla}_{n}h^j{_{k}}-2\bar{R}^n{_{k}}\bar{\nabla}_{l}h^j{_{n}}+R\bar{\nabla}_{l}h^j{_{k}}-h^j{_{n}}\bar{\nabla}_{l}\bar{R}^n{_{k}}\bigg ).
 \label{4}
 \end{aligned}
 \end{equation}
Using these results one arrives at
 \begin{equation}
\begin{aligned}
 \eta^{ikl}\,\bar{\nabla}_{k} (G{^{j}}{_{l}})_L= \frac{1}{2}\eta^{ikl}\,&\bigg (-2h\bar{\nabla}_{k}\bar{R}_{l}{^{j}}-\bar{R}_{l}{^{j}}\bar{\nabla}_{k}h
 -h^{j}{_{l}}\bar{\nabla}_{k}\bar{R}+3\bar{R}^{n j}\bar{\nabla}_{k}h_{n l}+3h_{n l}\bar{\nabla}_{k}\bar{R}^{n j}\\
&+\bar{R}_{nk}\bar{\nabla}^{j}h^{n}{_{l}}
-h^{n}{_{l}}\bar{\nabla}_{n}\bar{R}^{j}{_{k}}
 +\bar{R}^{n}{_{k}}\bar{\nabla}_{l}h^{j}{_{n}}
- \bar{R}_{kn}\bar{\nabla}^{n}h^{j}{_{l}}\\
&+\bar{\nabla}^{j}\bar{\nabla}_{n}\bar{\nabla}_{k}h_{l}{^{n}}
-\bar{\square}\bar{\nabla}_{k}h^j{_{l}}\bigg).
 \label{Linearresult}
 \end{aligned}
 \end{equation}
The $B$ term in (\ref{Linearcotton}) is easy to handle:
 \begin{equation}
  \eta^{ikl}\,(\Gamma^{j}{_{k n}})_L\bar{G}^{n}{_{l}}=\frac{1}{2}\eta^{ikl}\,\bar{R}^{n}{_{l}}\Big(\bar{\nabla}_{k}h_{n}{^{j}}
  +\bar{\nabla}_{n}h^{j}{_{k}}-\bar{\nabla}^{j}h_{k n}\Big).
  \label{6}
 \end{equation} 
Thus, collecting all these pieces, one arrives at the desired equation:
 \begin{equation}
\begin{aligned}
 2( C^{ij})_L=&-\frac{3h}{2}\,{\bar{C^{ij}}}-\frac{1}{2}\eta^{ikl}\,\bar{\square}\bar{\nabla}_{k}h^{j}{_{l}}+\frac{1}{2}\eta^{ikl}\,\bar{\nabla}^{j}\bar{\nabla}_{n}\bar{\nabla}_{k}h_{l}{^{n}}
+ \frac{3}{2}\eta^{ikl}\,\bar{\nabla}_{k}(\bar{R}^{n j}h_{n l})
-\frac{1}{2}\eta^{ikl}\,\bar{R}_{l}{^{j}}{\nabla}_{k}h\\
&-\frac{1}{2}\eta^{ikl}\,h^{n}{_{l}}\bar{\nabla}_{n}\bar{R}^{j}{_{k}}-\frac{1}{2}\eta^{ikl}\,h^{j}{_{l}}\bar{\nabla}_{k}\bar{R} 
+\eta^{ikl}\,\bar{R}_{nk}{\nabla}^{j}h^{n}{_{l}}+\eta^{ikl}\,\bar{R}_{l}{^{n}}{\nabla}_{n}h^{j}{_{k}}+i \leftrightarrow j.
 \label{cotlin}
 \end{aligned}
 \end{equation}
 
\section{Appendix C: Linearized Deformation of the Conformally flat background}
In this Appendix,  let us give a more explicit linearization of the Cotton flow equation about its conformally flat fixed point. Any metric of the form
\begin{equation}
ds^2= \Omega^2(x, y,z)\Big (dx^2+dy^2+dz^2\Big),
\end{equation}
with a smooth $ \Omega(x, y,z)>0$ is a critical point of the flow. Consider the most general perturbation about a given fixed point as
\begin{equation}
ds^2=\Omega^2 \Big ((1+a) \,dx^2+2b \, dx dy+2c \, dx dz+(1+f)\, dy^2+2 g dy \, dz+(1+h) \, dz^2 \Big),
\end{equation}
where all the functions depend on ${\bf x}$ and $t$ with the assumption that,  save $\Omega$,  all functions vanish at large distances at $t=0$.
At first order in the perturbation theory,  one computes the components of the Cotton tensor as 
\begin{equation}
\begin{aligned}
C_{11}&=\frac{1}{2 \Omega^3} \Big ( \partial^3_z b-\partial_y \partial^2_z c+\partial^2_y \partial_z b-\partial^3_y c-\partial_x \partial^2_z g-\partial_x \partial_y \partial_z f+ \partial_x \partial_y \partial_z h+\partial_x \partial^2_y g  \Big), \\
C_{12}&=\frac{1}{4 \Omega^3} \Big (- \partial^3_z a+\partial^3_z f-2 \partial_y  \partial^2_z g-\partial^2_y  \partial_z a+\partial^2_y  \partial_z h+2 \partial_x  \partial^2_z c+2 \partial_x  \partial^2_y c+ \partial^2_x  \partial_z f \\
&\hskip 1.5 cm-\partial^2_x  \partial_z h-2 \partial^2_x  \partial_y g \Big ),\\
C_{13}&=\frac{1}{4 \Omega^3} \Big ( \partial_y \partial^2_z a- \partial_y \partial^2_z f+2 \partial^2_y \partial_z g+ \partial^3_y a- \partial^3_y h- 2 \partial_x \partial^2_z b- 2 \partial_x \partial^2_y b+2\partial^2_x \partial_z g \\
& \hskip 1.5 cm+\partial^2_x \partial_y f- \partial^2_x \partial_y h \Big ), \\
C_{22}&=\frac{1}{2 \Omega^3} \Big ( -\partial^3_z b+\partial_y \partial^2_z c  +\partial_x \partial^2_z g+\partial_x \partial_y \partial_z a -\partial_x \partial_y \partial_z h-\partial^2_x \partial_z b -\partial^2_x \partial_y c +\partial^3_x g \Big ), \\
C_{23}&=\frac{1}{4 \Omega^3} \Big (2\partial_y \partial^2_z b-2\partial^2_y \partial_z c+\partial_x \partial^2_z a-\partial_x \partial^2_z f-\partial_x \partial^2_y a+\partial_x \partial^2_y h-2 \partial^2_x \partial_z c+2 \partial^2_x \partial_y b \\
&\hskip 1.5 cm-\partial^3_x f+ \partial^3_x h\Big),\\
C_{33}&=\frac{1}{2 \Omega^3} \Big (-\partial^2_y \partial_z b+\partial^3_y c-\partial_x \partial_y \partial_z a+\partial_x \partial_y \partial_z f-\partial_x \partial^2_y g+\partial^2_x \partial_z b + \partial^2_x \partial_y c-\partial^3_z g \Big).
\end{aligned}
\end{equation} 
Even though one can use the symmetries of the equation to reduce the system, this is still an unwieldy set. But our task is not to find the general solution rather to find the potentially unstable modes, hence let us assume for the  sake of simplicity that the perturbations depend on $z$ and $t$ but not on $x$ and $y$. Then one arrives at
\begin{equation}
\begin{aligned}
 &\partial_t (\Omega^2 a)=\frac{1}{2\Omega^3}\partial_z^3b, \hskip 1.cm \partial_t(\Omega^2f)=-\frac{1}{2\Omega^3}\partial_z^3b, \hskip 1.5cm \partial_t(\Omega^2 b)=\frac{1}{4\Omega^3}\partial_z^3(f-a), \\
&\partial_t (\Omega^2c)=0,  \hskip 2.2cm\partial_t(\Omega^2g)=0,\hskip 2.6cm \partial_t(\Omega^2h)=0.
\label{condition2}
\end{aligned}
\end{equation}
Further assuming $c(0, z)=h(0, z)=g(0, z)=0 $, the second line of  (\ref{condition2}) yields the vanishing of these functions for all $t$. From the first two equations of the first line, one finds $f(t, z)=-a(t,z)$.
Therefore, we have 
\begin{equation}
 \partial_t (\Omega^2 a)=\frac{1}{2\Omega^3}\partial_z^3b,\hskip 1.5cm \partial_t(\Omega^2 b)=-\frac{1}{2\Omega^3}\partial_z^3a,
\end{equation}
which yield a linearized, complex KdV-type equation with a variable coefficient
\begin{equation}
  \partial_t\Big (\Omega^2 (a\pm ib) \Big)=\mp\frac{i}{2\Omega^3}\partial_z^3(a\pm ib).
  \label{polarization1}
\end{equation}  
For flat space $\Omega=1$ and the Fourier transform of this equation yields the modes found in the text
\begin{equation}
 w({\bf p})=\pm \frac{i}{2}p^2 \lvert {\bf p} \rvert,
\end{equation}
with the $+$ mode giving the perturbative instability. We have not been able to solve (\ref{polarization1}) for generic $\Omega(x, y, z, t)$, but assuming $\Omega=\Omega(x, y)$ one finds the dispersion relation 
\begin{equation}
 w({\bf p})=\pm \frac{i}{2 \Omega^5(x, y)}p^2 \lvert {\bf p} \rvert,
\end{equation}
hence the unstable mode survives for this type of conformally flat background backgrounds. Therefore, it is clear that among the conformally flat fixed points of the flow, some metrics are saddle points rather than being the minima. It is an outstanding problem to find the minima of the Cotton flow. 

\section{Appendix D: Cotton tensor under an arbitrary flow and under Ricci flow}
Our computation in Appendix B allows us to compute the flow of the Cotton tensor under an arbitrary geometric flow defined as
\begin{equation}
\partial_tg_{ij}=E_{ij},
\end{equation}
where $E_{ij}$ is a symmetric tensor. With the help of equation (\ref{cotlin}), one gets
  \begin{equation}
\begin{aligned}
 2 \partial_tC^{ij}=&-\frac{3E}{2}\,{\bar{C^{ij}}}-\frac{1}{2}\eta^{ikl}\,\bar{\square}\bar{\nabla}_{k}E^{j}{_{l}}+\frac{1}{2}\eta^{ikl}\,\bar{\nabla}^{j}\bar{\nabla}_{\lambda}\bar{\nabla}_{k}E_{l}{^{\lambda}}
+ \frac{3}{2}\eta^{ikl}\,\bar{\nabla}_{k}(\bar{R}^{\alpha j}E_{\alpha l})
-\frac{1}{2}\eta^{ikl}\,\bar{R}_{l}{^{j}}{\nabla}_{k}E\\
&-\frac{1}{2}\eta^{ikl}\,E^{n}{_{l}}\bar{\nabla}_{n}\bar{R}^{j}{_{k}}-\frac{1}{2}\eta^{ikl}\,E^{j}{_{l}}\bar{\nabla}_{k}\bar{R} 
+\eta^{ikl}\,\bar{R}_{nk}{\nabla}^{j}E^{n}{_{l}}+\eta^{ikl}\,\bar{R}_{l}{^{\alpha}}{\nabla}_{\alpha}E^{j}{_{k}}+i \leftrightarrow j.
\label{cotric}
 \end{aligned}
 \end{equation}
It is interesting to see how the Cotton tensor behaves under the normalized and unnormalized Ricci flow which we give below. 
 Cotton tensor under unnormalized Ricci flow $\partial_tg_{ij}=-2R_{ij}$  flows as 
\begin{equation}
  2 \partial_t C^{ij}=\square C^{ij}+4R C^{ij}-6R^{kj}C^i{_k}+2g^{ij}R_{nk}C^{nk}+\eta^{ikl}R_l{^j}\nabla_k R-2\eta^{ikl}R_{nl}\nabla_k R^{nj}+i \leftrightarrow j.
\end{equation}
 Cotton tensor under normalized Ricci flow $\partial_tg_{ij}=-2(R_{ij}-\frac{1}{3}g_{ij}R)$ flows as
\begin{equation}
 2 \partial_t C^{ij}=\square C^{ij}+\frac{7}{3} R C^{ij}-6R^{kj}C^i{_k}+2g^{ij}R_{nk}C^{nk}+\eta^{ikl}R_l{^j}\nabla_k R-2\eta^{ikl}R_{nl}\nabla_k R^{nj}+i \leftrightarrow j.
\end{equation}
It is nice to see that both equations yield a parabolic flow of the Cotton tensor and hence amenable to maximum principle.  

\section{\label{ackno} Acknowledgments}

We would like to thank O. Kisisel for useful discussions. The work of B.T. and S.D. is partially supported by the TUBITAK Grant No.113F155. E.K. is partially supported by the TUBITAK PhD Scholarship.

\end{document}